\documentclass[pre,showpacs,preprint,endfloats*]{revtex4}
\usepackage{amsmath}
\usepackage{amsfonts}
\usepackage[final]{graphicx}

\def\complex{\mathbb{C}}
\def\real{\mathbb{R}}
\def\sphere{\mathbb{S}}
\DeclareMathOperator\Real{Re}
\DeclareMathOperator\Imag{Im}

\begin{document}
\title{Completely monotone solutions of the mode-coupling theory
       for mixtures}
\date{\today}
\author{T.~Franosch}
\affiliation{Hahn-Meitner-Institut, Abteilung Theoretische Physik,
             Glienicker Str.~100, 14109 Berlin, Germany}
\author{Th.~Voigtmann}
\affiliation{Physik-Department, Technische Universit\"at M\"unchen,
             James-Franck-Str., 85747 Garching, Germany}

\begin{abstract}
We establish that a mode-coupling approximation for the dynamics of
multi-component systems obeying Smoluchowski dynamics
preserves a subtle yet fundamental property: the matrices of partial density
correlation functions are completely monotone, i.e.\ they can exactly be written
as superpositions of decaying exponentials only. This statement holds,
no matter what further approximations are needed to calculate the theory's
coupling parameters. The long-time limit of these functions fulfills a maximum
property, and an iteration scheme for its numerical determination is given.
We also show the existence of a unique solution to the equations of motion
for which power series both for short times and small frequencies exist,
the latter except at special points where ergodic-to-nonergodic transitions
occur. These transitions are bifurcations that are proven to be of the
cuspoid family.
\end{abstract}
\pacs{82.70.Dd, 05.40.-a, 02.10.Sp}

\maketitle

\section{Introduction}

Density correlation functions are convenient tools to characterize the
dynamics of liquids or disordered systems. They can be measured
in experiment, e.g.\ by inelastic neutron scattering, dynamic light scattering
in colloidal systems, or can be determined
from computer simluation techniques.
On the other hand, one can calculate them from theory, but in the case of
strongly interacting systems, one typically has to invoke certain
approximations in order to obtain the desired results.
There are, however, some general properties of such correlation functions,
directly related to the time-evolution operator of the system. It is
a nontrivial point to show that all approximations involved in deriving
a theory's equations of motion preserve these general properties.

One example are colloidal suspensions obeying Smoluchowski
dynamics. One knows from general grounds that in such systems
the matrices of partial correlation functions are completely monotone,
i.e.\ they can be written as a superposition of decaying exponentials only.
An approximative theory calculating such quantities should aim to
reproduce these properties, since they are direct consequences of the
structure of the time-evolution operator. However, the concept of
complete monotonicity is quite subtle, and it is therefore likely that
a given approximation prevents the approximative solutions from sharing this
feature with the complete solution.

At high densities, colloidal systems are known to undergo
glassy dynamics, provided crystallization can be suppressed for a
sufficiently long time. In these cases the so-called mode-coupling theory
of the glass transition (MCT) has been successful in describing much of the
experimental facts. For one-component, i.e.\ monodisperse, systems the
theory has been proven \cite{Goetze1995b} to give results for the
density correlation functions, which indeed reproduce the above mentioned
features. However, little is known about multi-component, or polydisperse,
mixtures. Recently, glassy dynamics in a binary colloidal suspension has been
studied \cite{Williams2001b}, challenging detailed comparisons of the MCT
for mixtures with experiment.

MCT tries to describe the motion of particles on a
microscopic scale by an approximation of the frequency- and
wavenumber-dependent viscosity in terms of density-fluctuation products.
Such an approach comes about naturally if one considers the potential stresses
to be built up by the density fluctuations of the liquid itself. The resulting
equations contain a feedback mechanism induced by the slowing down of
relaxations due to the so-called cage effect.
It has been shown that these equations allow for an ideal glass state, given
the interactions are strong enough, e.g.\ at high densities. The ideal glass
is characterized by a nonvanishing long-time limit of density correlation
functions which corresponds to an elastic scattering contribution in the
dynamical structure factor.
In the ``phase diagram'', there occur critical manifolds, referred to as
glass-transition singularities, that separate ergodic liquid from
ideal glass states and can also extend into the glass state.
Upon approaching a singularity point, the long-time
limit of the density correlators, sometimes referred to as the glass form
factor or the Edwards-Anderson parameter of
the system, changes discontinuously. This holds for both liquid-glass
transitions, where the change is from zero to some nonzero value, as well as
glass-glass transitions.

Close to the transition, analytic formulas providing asymptotic solutions of
the MCT equations of motion have been derived. There occur two time fractals
with nontrivial exponents, and two diverging time scales, accompanied by
corresponding scaling laws. On these time scales the dynamics of the
glass former is also referred to as structural relaxation.
The asymptotic predictions as well as numerical solutions
of the full MCT equations for model systems have been extensively tested
against experimental data; for a detailed discussion of the glass-transition
scenario, the reader is referred to a recent review \cite{Goetze1999}.

The aim of this paper is to generalize
from the case of one-component systems to that of multi-component mixtures
proofs of the basic properties of MCT solutions.
In particular, we show that the structural relaxation can be represented as
a continuous superposition of decaying exponentials, i.e.\ that the density
correlation functions indeed are completely monotone. Furthermore, the
long-time limits
can be obtained by a simple iteration procedure that does not involve the
solution of the complete dynamical equations.
An investigation of this iteration brings out glass transitions in general
to be bifurcations of the cuspoid type.
In addition, a short-time expansion of the density correlation functions
is demonstrated to be convergent for short times, and similarly, a power
series for small frequencies is shown to exist.

The paper is organized as follows:
In Sec.~\ref{sec.eom}, we will introduce the equations of motion for the
mixtures considered, together with some basic properties and the MCT
approximation. Section~\ref{compmon} presents a proof that the MCT equations
of motion have a uniquely determined solution which is completely monotone.
In Sec.~\ref{dwf}, the long-time limit of these solutions will be
discussed, and in Sec.~\ref{powerseries}, the existence of power series
solutions both in the time and frequency domain is shown, the latter by
proving first that all moments of the correlation functions are finite.
Section~\ref{conclusio} offers some conclusions.

\section{Equations of Motion}\label{sec.eom}

\subsection{General Properties}

We consider a classical system enlcosed in a box of volume $V$ with a total
number of 
particles $N = \sum_{\nu=1}^s N_\nu$, consisting of $s$ different species with
number concentrations $x_\nu = N_\nu / N$.
The particles are supposed to be structureless, i.e.\ they do
not possess any internal degrees of freedom, and are thus fully described by
their positions, momenta, and species index. The variables
\begin{equation}
  n_\nu(\vec q) = N^{-1/2}\sum_{i=1}^{N_\nu} \exp(i\vec q\cdot\vec r_i^\nu)
\end{equation}
are then the fluctuating densities of species $\nu$ to wave vector $\vec q$.
Here, $\vec r_i^\nu$ denotes the position of the $i$-th particle of species
$\nu$. The simplest statistical information on structural dynamics
that can be extracted from a
multi-component supercooled liquid is the matrix of density correlation
functions
\begin{equation} \label{density}
  \Phi_{\mu\nu}(q,t) =
  \left\langle n_\mu(\vec q,t)|n_\nu(\vec q)\right\rangle\,.
\end{equation}
Here, the brackets $\langle\cdot|\cdot\rangle$ denote the Kubo product with
$\langle A|B\rangle = \langle\delta A^*\delta B\rangle$, where
$\delta A = A - \langle A \rangle$, and $\langle\dots\rangle$ indicates
canonical averaging. Since $\Phi_{\mu\nu}(q,t)$ is the spatial Fourier
transform of a function that is real, translational-invariant and isotropic,
it is itself real and depends only on the magnitude of the wave vector
$q = |\vec q|$. The time evolution is given by
$n_\mu(\vec q,t) = \exp[i{\mathcal S}t] n_\mu(\vec q)$.
For a liquid obeying Newtonian dynamics, the operator $\mathcal S$ is just the
Liouville operator, which is Hermitian with respect to the Kubo product.
In this case, time inversion symmetry implies the density correlation matrix
to be symmetric with respect to interchange of the species indices.

Let us focus on colloidal liquids, and $\mathcal S$ denote the Smoluchowski
operator. There, time inversion symmetry is broken explicitly, but still
the symmetry $\Phi_{\mu\nu}(q,t)=\Phi_{\nu\mu}(q,t)$ holds, although this
has to be proven seperately \cite{Naegele1996}.

One then has the spectral decomposition
$\exp[i{\mathcal S}t]=\int\exp(-\gamma t)dP_\gamma$, with eigenvalues $\gamma$
fulfilling $\gamma\geq0$, and $P_\gamma$ denoting the projector onto the
corresponding linear subspace. This immediately leads to the following
representation:
\begin{equation}\label{represent}
  \Phi_{\mu\nu}(q,t) = \int e^{-\gamma t} da_{\mu\nu,q}(\gamma)\,,
\end{equation}
where the measure $a$ is concentrated on the nonnegative real axis,
is symmetric in $\mu,\nu$, and positve:
$da_{\mu\nu,q}(\gamma)\succeq0$,
i.e.\ for any set of complex numbers $y_\nu$, $\nu = 1,\dots,n$,
the measure $y_\mu^*\,da_{\mu\nu,q}(\gamma)\,y_\nu$ is positive (summation over
repeated indices is implied here and in the following). A function having
these properties is called completely monotone. In particular,
$\Phi_{\mu\nu}(q,t)$ is a positive definite matrix for all times, and for all
$l$ the time derivatives $(-1)^l \partial_t^l \Phi_{\mu\nu}(q,t)$ are
positive definite. The equivalence of these formulations is the result of
the Bernstein theorem \cite{Gripenberg1990,Widder1946}.

Let us also introduce the Laplace transform
$\Phi_{\mu\nu}(q,z) = i \int_0^\infty e^{i z t} \Phi_{\mu\nu}(q,t)\,dt$.
Then the representation Eq.~(\ref{represent}) shows that
\begin{equation}\label{laplace}
  \Phi_{\mu\nu}(q,z) = \int \frac{-1}{z + i \gamma}\,da_{\mu\nu,q}(\gamma)  
\end{equation}
is
(i) analytic for $z\in\complex\setminus i\real^-$,
(ii) obeys $\Phi_{\mu\nu}(q,z)^*=\Phi_{\mu\nu}(q,-z^*)$,
(iii) $\lim_{z\to i\infty} \Phi_{\mu\nu}(q,z)=0$, and
(iv) $\Real \Phi_{\mu\nu}(q,z)\succeq0$ for $\Real z<0$.
In reverse, these four properties are enough in order to guarantee
a representation in the form of Eq.~(\ref{represent})
\cite[Section~5, Theorem~2.6]{Gripenberg1990}.
The spectrum $\Phi_{\mu\nu}''(q,\omega)=\Imag\Phi_{\mu\nu}(q,z=\omega+i0)$
then is a superposition of Lorentzians
\begin{equation}
  \Phi_{\mu\nu}''(q,\omega)
  =\int\frac{\gamma}{\omega^2+\gamma^2}\,da_{\mu\nu,q}(\gamma)\,,
\end{equation}
which is positive as is already implied by the passivity of the system.
One also has that the long-time limit of the correlators exists. If this
quantity is nonvanishing,
\begin{equation}
  F_{\mu\nu}(q) = \Phi_{\mu\nu}(q,t\to \infty) \neq 0\,,
\end{equation}
it is called the glass form factor or nonergodicity parameter,
and the spectrum exhibits an elastic contribution
$\pi F_{\mu\nu}(q)\delta(\omega)$.
Passivity requires $F_{\mu\nu}(q)\succeq0$ which is consistent
with Eq.~(\ref{represent}). 

Let us stress again that the above properties of density autocorrelation
functions are direct consequences of the eigenvalue spectrum of the
Smoluchowski operator; in general,
they hold for any system whose time-evolution operator has nonnegative, real
eigenvalues only.

\subsection{Mode-Coupling Theory}

Mode-coupling theory starts from the formally exact representation of the
density correlation matrix in terms of a memory kernel matrix.
In the Laplace domain this results in the matrix equation
\begin{equation}\label{eom}
  \Phi(q,z) = -\left[z S^{-1}(q)
  -S^{-1}(q)\left[i\tau(q)+M(q,z)\right]^{-1}S^{-1}(q)
  \right]^{-1}\,.
\end{equation} 
Since $M(q,z)\to0$ for $z\to\infty$, one can identify the matrices
$S$ and $\tau$
with the short time expansion of the time density correlation function,
$\Phi(q,t)=S(q)-\tau^{-1}(q) t + {\mathcal O}(t^2)$.
The matrix $S(q)$ is called the structure factor, and
from the definition Eq.~(\ref{density}) one checks that for every $q$ it is 
symmetric, real and positive definite. The same properties hold for the matrix
$\tau(q)$ characterizing the initial decay of $\Phi(q)$.
We shall throughout this paper discuss the above equation for $q\neq0$
and therefore assume both $S(q)$ and $\tau(q)$ to be invertible,
which is the generic case as long as all number concentrations are
nonvanishing.

The Zwanzig-Mori formalism gives an explcit expression for the memory
matrix in terms of so-called fluctuating forces. We shall only be concerned
with its general structure, which in the time domain is
\begin{equation}
  M_{\mu\nu}(q,t) = \langle X_\mu(\vec q) |
  {\mathcal R}'(t) X_\nu(\vec q) \rangle\,.
\end{equation} 
Explicit expressions for the fluctuating forces $X_\mu(\vec q)$ and 
the reduced resolvent ${\mathcal R}'(t)$ can be worked out
\cite{Boon1980,Hansen1986}.
The mode-coupling approximation consists of projecting the fluctuating forces
onto pair modes $n_{\sigma}(\vec k)n_{\tau}(\vec p)$ and
factorizing four-particle correlations
\cite{Bengtzelius1984,Goetze1991b},
\begin{equation} 
\langle n_\sigma(\vec k)n_\tau(\vec p) | {\mathcal R}'(t) 
  n_{\sigma'}(\vec k') n_{\tau'}(\vec p') \rangle
\approx
  \Phi_{\sigma\sigma'}(\vec k,t)\Phi_{\tau\tau'}(\vec p,t)\delta_{\vec k\vec k'}
  \delta_{\vec p\vec p'} \,.
\end{equation}
In order to avoid overcounting we restrict the pair modes
to $\vec k>\vec p$, $\vec k'>\vec p'$ with some order relation.  
In particular, for $t=0$ this implies an approximate normalization and suggests
to introduce an approximate projector
\begin{equation}
  {\mathcal P}_2\approx\sum_{\vec k>\vec p}
   \left| n_\sigma(\vec k)n_\tau(\vec p)\right\rangle 
  S_{\sigma\sigma'}^{-1}(k)S_{\tau\tau'}^{-1}(p)
   \left\langle n_{\sigma'}(\vec k)n_{\tau'}(\vec p)\right| \,.
\end{equation}
With this projector, the MCT approximation is \cite{Goetze1987b,Fuchs1993}
\begin{align} 
 M^{\text{MCT}}_{\mu\nu}(q,t)&=
  \sum_{\vec k>\vec p}
  \langle X_\mu(\vec q) | n_{\sigma'}(\vec k)n_{\tau'}(\vec p) \rangle 
  S_{\sigma'\sigma}^{-1}(k)S_{\tau'\tau}^{-1}(p) \times \nonumber \\
  &\qquad
  \Phi_{\sigma\bar\sigma}(k,t)\Phi_{\tau\bar\tau}(p,t)
  S_{\bar\sigma\bar\sigma'}^{-1}(k) S_{\bar\tau\bar\tau'}^{-1}(p)
  \langle n_{\bar\sigma'}(\vec k)n_{\bar\tau'}(\vec p) | X_\nu(\vec q)
  \rangle \,.
\end{align}
It is useful to write this in a more transparent form by introducing
super-indices $s=(\sigma,\tau)$, $\bar s=(\bar\sigma,\bar\tau)$:
\begin{equation} \label{MCT}
 M_{\mu\nu}(q,t)=
  \sum_{\vec k>\vec p} V_{\mu s}(\vec q,\vec k\vec p) 
  [\Phi(k,t)\otimes\Phi(p,t)]_{s\bar s}
  V_{\nu\bar s}(\vec q,\vec k\vec p)^*\,,
\end{equation}
where $\otimes$ denotes the tensor product in the space of species indices,
and the `vertex' reads
\begin{equation} \label{vertex}
  V_{\mu,s=(\sigma\tau)}(\vec q,\vec k\vec p) = 
  \langle X_\mu(\vec q) | n_{\sigma'}(\vec k)n_{\tau'}(\vec p) \rangle 
  S_{\sigma'\sigma}^{-1}(k)S_{\tau'\tau}^{-1}(p)\,.
\end{equation}
Since the tensor product of positive definite matrices again represents a
positive definite matrix in the corresponding product space, it is clear from
Eq.~(\ref{MCT}) that $M_{\mu\nu}(q,t)$ is positive definite, provided that the
density correlation matrix $\Phi_{\mu\nu}(q,t)$ is positive definite for all
wave vectors. More generally, if we write
\begin{equation} \label{functional}
  {\mathcal F}_{q,\mu\nu}[F,G]
  =\frac12\sum_{\vec k>\vec p}
  V_{\mu s}\left[F\otimes G
  +G\otimes F\right]_{s\bar s}V_{\nu\bar s}^*\,,
\end{equation}
we see that ${\mathcal F}[F,G]$ is symmetric in $F$ and $G$, and is positive
definite for every $q$, provided both $F\succeq0$ and $G\succeq0$ for every
$k$ and $p$. In particular, we have $M(q,t)={\mathcal F}_q[\Phi,\Phi]$.

Let us mention that the vertex can be evaluated and expressed in terms of the
static structure factor matrix and the three-particle static correlation
functions. Usually the structure factor is known only approximately.
Knowledge of triple correlations is often lacking entirely, although in
principle they can be determined from computer simulation
\cite{Sciortino2001}.
However, the property of $M_{\mu\nu}(q,t)$ being a positive definite quantity
is a direct consequence of the MCT approximation structure, and
is indeed independent of the approximations made in order to evaluate the
vertex in Eq.~(\ref{vertex}). 
Note in addition that one can in principle include a regular contribution
to the memory kernel, $M^{\text{reg}}(q,z)$, accounting for transient dynamics
not captured in the MCT approximation, e.g.\ hydrodynamic interactions and the
like. Under the assumption of such a term being completely monotone,
the following discussion remains valid.
That there indeed exists a completely monotone solution to Eqs.~(\ref{eom})
and (\ref{MCT}) shall be proven in the following section.

\section{Completely Monotone Solutions}\label{compmon}

\subsection{Complete Monotonicity}

We denote the space of $s\times s$ matrices, where $s$ is the number of
species, by $\mathcal A$. It is clear that $\mathcal A$, equipped with standard
matrix multiplication and Hermitean scalar product, is indeed a
$C^\star$ algebra. For elements $a_q\in{\mathcal A}$, we form
vectors $a=(a_q)_{q=1,\dots M}\in{\mathcal A}^M$. One easily checks that
${\mathcal A}^M$ with all matrix operators over $\mathcal A$ defined
elementwise in $q$, and equipped with the maximum norm $\|a\|=\max_q\|a_q\|$,
can again be turned into a $C^\star$ algebra. An element $a\in{\mathcal A}^M$
shall be called positive, $a\succeq0$, if $a_q\succeq0$ for every $q$;
similarly, we use $a\succ0$, or $a\succeq b$, the latter meaning $a-b\succeq0$.
Note that the norm preserves ordering, i.e.\ for $a\succeq b$ we also have
$\|a\|\ge\|b\|$.

In the following, we shall assume wave vectors to be discretized to some
finite set $q=1,\dots M$, such that all matrices appearing in the equations
of motion are elements of ${\mathcal A}^M$.

Now assume $\Phi^{(n)}(t)$ to be completely monotone. It is then clear from
Eq.~(\ref{MCT}) that $M^{(n)}(t)$ inherits this property. In particular, its
Laplace transform has the properties (i) to (iv) of Eq.~(\ref{laplace}).
But then $\Phi^{(n+1)}(z)$ as defined by
\begin{equation}\label{laplace-iteration}
  \Phi^{(n+1)}(z)=-\left[zS^{-1}-S^{-1}\left[i\tau+M^{(n)}(z)\right]^{-1}S^{-1}
  \right]^{-1}
\end{equation}
again fulfills properties (i) to (iv). This is easily checked for (i) to (iii).
Property (iv) can be shown in two steps. First, define $K(z)$ by
\begin{equation} \left[i\tau+M^{(n)}(z)\right]K(z)=-1\,.
\end{equation}
One then has
\begin{subequations}
\begin{eqnarray}
  (\Real M(z))(\Real K(z)) -\left(\tau+\Imag M(z)\right) (\Imag K(z))&=&-1\,,\\
  (\Real M(z))(\Imag K(z)) +\left(\tau+\Imag M(z)\right) (\Real K(z))&=&0\,.
\end{eqnarray}
\end{subequations}
Using the second equation, one can eliminate $\Imag K(z)$ in the first and
find for $\Real z<0$, where $\Real M(z)\succeq0$, that $\Real K(z)\preceq0$.
But we have
\begin{equation} \left[zS^{-1}+S^{-1}K(z)S^{-1}\right]\Phi(z)=-1\,,
\end{equation}
and along the same lines one derives with $\Real K(z)\preceq0$ for $\Real z<0$
the desired result, $\Real\Phi(z)\succeq0$.

Thus, Eq.~(\ref{laplace-iteration}) together with some completely monotone
starting point, $\Phi^{(0)}(t)=\exp[-(S\tau)^{-1}t]S$, say,
defines a sequence of completely monotone functions $\Phi^{(n)}(t)$,
normalized to $\Phi(t=0)=S$.
Let us complete the proof of existence of
a uniquely determined, completely monotone solution to Eqs.~(\ref{eom})
and (\ref{MCT}) by
showing that the thus constructed sequence converges uniformly to some
$\Phi(t)$.

\subsection{Existence and Uniqueness}

The equation of motion in the time domain can be obtained from Eq.~(\ref{eom})
which yields
\begin{equation} \label{time}
  \tau\dot\Phi(t) + S^{-1}\Phi(t) 
  + (M\ast\dot\Phi)(t)=0\,.
\end{equation}
Here, $(f\ast g)(t)=\int_0^t f(t-t')g(t')dt'$ denotes the time-domain
convolution.
The density correlation matrix is subjected to the initial condition
$\Phi(t=0)=S$, and the memory kernel $M(t)$ is given in terms of the
density correlators, Eq.~(\ref{MCT}). 

This equation of motion can be rewritten as an integral equation similar to
the Picard equation,
\begin{equation}
  \tau\Phi(t)=\tau S +\int_0^t \left[ M(t')S-S^{-1}
  \Phi(t') -M(t')\Phi(t-t') \right]\,dt'\,,
\end{equation}
such that the standard proof of local existence and uniqueness can be
applied. In particular, the Picard iteration corresponding to the
Laplace-domain iteration defined in the previous subsection is
\begin{equation} 
 \Phi^{(n+1)}(t)
  =S+\int_0^t{\mathcal K}
  \left[\Phi^{(n)}(t'),\Phi^{(n)}(t-t'),
        \Phi^{(n+1)}(t')\right]\,dt'\,,
\end{equation}
where
\begin{equation}
  {\mathcal K}[x,y,z]=\tau^{-1}\left({\mathcal F}[x]S
  -\left(S^{-1}+{\mathcal F}[y]\right)z\right)\,.
\end{equation}
The convergence of this iteration is proven as in the one-component case
\cite{Goetze1995b}, using a Lipschitz constant $L$ for $\mathcal K$. If
we restrict $t$ to some finite time interval, $0\le t\le T$, and the vertices
$V$ to some finite closed domain, we get, since $\|\Phi^{(n)}(t)\|\le\| S\|$
ensures a finite closed domain for $x$, $y$, $z$,
\begin{equation}
  \|{\mathcal K}(x_1,y_1,z_1)-{\mathcal K}(x_2,y_2,z_2)\|
  \le L\left(\|x_1-x_2\|+\|y_1-y_2\|+\|z_1-z_2\|\right)\,.
\end{equation}
This allows to construct a sequence $X_n(t)=\|\Phi^{(n+1)}(t)-\Phi^{(n)}(t)\|
/\|S\|$ obeying the inequalities $X_n(t)\le1$,
\begin{equation} X_n(t)\le L\int_0^t(2X_{n-1}(t')+X_n(t'))dt'\,, \end{equation}
and from there, the proof of Ref.~\cite{Goetze1995b} applies.
Thus there exists for each
fixed finite $T$ a unique solution in the interval $0\le t\le T<\infty$.
Furthermore, this solution depends smoothly on
the vertices $V$ as control parameters for any fixed finite time interval.
Note that this theorem cannot be extended to infinite time intervals.

Together with the preceding paragraph, we have proven the existence of a
unique solution to Eq.~(\ref{eom}) that it is the limit of a convergent
sequence of completely monotone functions. It is therefore itself completely
monotone, due to the continuity theorem for Laplace transforms
\cite{Feller1971b}.

\section{Glass Form Factors}\label{dwf}

In this section we shall prove that the glass form factor
$F=\Phi(t\to\infty)$ can be obtained without solving the integro-differential
equation (\ref{time}).

Since the Laplace-transform exhibits a pole at zero frequency,
$\Phi(z)=-F/z+\{\text{smooth}\}$,
Eq.~(\ref{eom}) implies that the form factor is a solution of
\begin{equation} \label{DW}
  F=S-\left[S^{-1}+N\right]^{-1}\,.
\end{equation}
Here, $N_{\mu\nu}(q)=M_{\mu\nu}(q,t\to \infty)$ denotes the long time limit of
the memory kernel matrix which is, according to Eqs.~(\ref{MCT}) and
(\ref{functional}), a quadratic functional of the form factor,
$N={\mathcal F}[F,F]$.

In general the coupled equations (\ref{DW}) and (\ref{functional}) have several
solutions, e.g.\ $F=0$ trivially satisfies the equations.
Since we have shown in the last section that the solution of the mode-coupling 
equation is symmetric and completely monotone, one can restrict the search on
the positive symmetric solutions of the above equations. In this space,
the solution corresponding to the glass form factor shall be shown to be
maximal with respect to the semi-ordering $\succeq$ defined on ${\mathcal A}^M$.

\subsection{Maximum Fixed Point}

The mode-coupling functional preserves the semi-ordering. Since
${\mathcal F}[F,G]\succeq0$ is satisfied for $F,G\succeq0$, and ${\mathcal F}$
is symmetric in $F$ and $G$, we have $N[F]-N[G]={\mathcal F}[F+G,F-G]\succeq0$,
if $F-G\succeq0$. Thus one finds $N[F]\succeq N[G]$ if $F\succeq G$.
It is easy to see that inversion reverses the semi-ordering, i.e.\ for
$F\succeq G\succ0$ one has $G^{-1}-F^{-1}\succeq0$.

Equations (\ref{DW}) and (\ref{functional}) suggest to introduce a continuous
mapping for a set of positve symmetric matrices $F$ by
\begin{equation}\label{mapping}
  {\mathcal I}[F] = S-\left[S^{-1}+N[F]\right]^{-1}\,.
\end{equation}
It is clear from the preceeding paragraphs that ${\mathcal I}[F]$
is again positive and symmetric and preserves the semi-ordering,
${\mathcal I}[F]\succeq{\mathcal I}[G]$ if $F\succeq G$. By induction one shows
that the sequence $F^{(n+1)}={\mathcal I}[F^{(n)}]$, $n=0,1,\dots$, starting
with $F^{(0)}=S\succ0$ is monotone and bounded, $S\succ F^{(n)}\succeq
F^{(n+1)}\succeq0$, $n=1,2,\dots$, and thus converges to some fixed point
$F^*\succeq0$ which is a solution of Eqs.~(\ref{DW}) and (\ref{functional}). 

Suppose now there is some positive definite, symmetric fixed point
$F^{**}$. If we introduce the mapping $F\mapsto\tilde F$ by $F=F^{**}+\tilde F$,
this maps $F=F^{**}$ to $\tilde F=0$, and $S\succ0$ to $\tilde S=S-F^{**}
\succ0$.
The mapping is covariant in the sense that $\tilde F
=\tilde{\mathcal I}[\tilde F]$ holds iff $F={\mathcal I}[F]$, provided one sets
$\tilde{\mathcal I}[\tilde F]=\tilde S-[\tilde S^{-1}+\tilde N[\tilde F]]^{-1}$
with
\begin{equation}
  \tilde N[\tilde F]=N[F]-N[F^{**}]\,.
\end{equation}
It is clear that the mapping $\tilde{\mathcal I}[\tilde F]$ exhibits
the properties of $\mathcal I$ discussed above. Thus the sequence
$\tilde F^{(n)}=F^{(n)}-F^{**}$ with
$F^{(n)}$ as above converges to some positive definite fixed point
$\tilde F^*$. By continuity of all maps involved,
$\tilde F^*=F^*-F^{**}\succeq0$, and thus
$F^*\succeq F^{**}$ for any fixed point $F^{**}$.
We can summarize that $F^*$ is a maximum fixed point
in the sense that is is larger than all other positive definite, symmetric
solutions of Eqs.~(\ref{DW}) and (\ref{functional}) with respect to the
semi-ordering introduced above. The iteration scheme defined by
${\mathcal I}$ converges to this maximum fixed point, provided the iteration
is started with the upper limit $S$.

\subsection{Uniqueness and Eigenvalue}

Denote by $\psi$ the linearization of ${\mathcal I}$ and thus $S{\mathcal F}S$.
It is clear that $\psi$ is a positive linear map on ${\mathcal A}^M$ in the
sense that $\psi[f]\succeq0$ for all $f\succeq0$.
For the one-component case considered in Ref.~\cite{Goetze1995b}, $\psi$
corresponds to a positive matrix $A_{qp}$ in the sense that $A_{qp}\ge0$
for all $q$ and $p$. From the physical picture of the MCT approximation it
is reasonable to assume that $A$ has no invariant subspaces, and thus is
an irreducible matrix. Then the Perron-Frobenius theorem \cite{Gantmacher1974b}
can be invoked to prove the existence of a non-degenerate, positive eigenvector
$z$ corresponding to the spectral radius of $A$.

In the present case, the generalization of this result is guaranteed since
the equivalent of the Perron-Frobenius theorem holds for positive linear maps
on $C^\star$ algebras \cite{Evans1978}. The physical interpretation again
leads to the assumption that $\psi$ is irreducible. Then in particular,
the mapping $\psi$
has a non-degenerate maximum eigenvalue $r$, to which there corresponds a
uniquely determined eigenvector $z\succ0$. For all other eigenvalues $\lambda$,
there holds $|\lambda|\le r$.
For completeness, a proof of the general Perron-Frobenius results as far as
needed here is sketched in Appendix~\ref{perron}.

If we suppose $r=1+\delta$ with some $\delta>0$, we have, after the
transformation ${\mathcal I}\mapsto\tilde{\mathcal I}$ as defined above,
\begin{equation}
  \tilde{\mathcal F}[\xi z]\succeq \psi[\xi z]=(1+\delta)\xi z
\end{equation}
with some real $\xi>0$. If we set $\hat F^{(0)}=\xi z$ and
define a sequence $\hat F^{(n)}$ by $\hat F^{(n+1)}
=\hat{\mathcal I}[\hat F^{(n)}]=S^{-1}\tilde{\mathcal I}[\hat F^{(n)}]S^{-1}$
for $n=1,2,\dots$, we have
$S-S\hat F^{(1)}S\preceq[S^{-1}+(1+\delta)\hat F^{(0)}]^{-1}$.
But there exists some $\varepsilon>0$ such that
$S^{-1}+(1+\delta)\hat F^{(0)}-S^{-1}(S^{-1}-\hat F^{(0)})^{-1}S^{-1}
=S^{-1}+(1+\delta)\hat F^{(0)}-S^{-1}(S+S\hat F^{(0)}S+\dots)S^{-1}
=\delta\cdot\hat F^{(0)}+{\mathcal O}(\xi^2)\succ0$ for all
$0<\xi\le\varepsilon$, and one gets $\hat F^{(1)}\succ\hat F^{(0)}\succ0$.
Since $\hat{\mathcal I}$
inherits the properties of $\tilde{\mathcal I}$, it follows that
$S^{-1}\succ\hat{\mathcal I}[\hat F^{(n+1)}]\succeq
\hat{\mathcal I}[\hat F^{(n)}]\succ0$. Thus the sequence $\hat F^{(n)}$,
$n=0,1,\dots$, is monotone and bounded, and by
continuity of $\hat{\mathcal I}$ converges to some fixed point
$\hat F^{\#}$. If we now choose $\tilde{\mathcal I}$ such
that $F^{**}=F^*$ is the maximum fixed point of ${\mathcal I}$,
$\hat F^{\#}\succ0$ implies the existence of some fixed point
$F^\#\succ F^*$ of ${\mathcal I}$. Thus, by
contradiction, $\delta>0$ cannot be possible, and we conclude $r\le1$,
i.e.\ the maximum eigenvalue of $\psi$ is bounded by unity.
The value of $r$ depends on the control parameters $V$, and thus one
distinguishes regular points, $r<1$, from the `critical' manifold, where
$r=1$.
Let us note that $F^*$ defined as the maximum fixed point of Eq.~(\ref{DW})
exhibits bifurcation at critical points, identified within MCT as the
ideal glass transition singularities. The non-degeneracy of $r$ implies that
MCT describes glass transitions in multi-component colloidal systems as
bifurcations of the $A_\ell$ type, according to the classification of
Arnol'd \cite{Arnold1975}. This in turn ensures that asymptotic solutions
can be worked out in the same spirit as for one-component systems, with
the common case being the dynamics close to a fold ($A_2$) bifurcation
\cite{Franosch1997}.

\subsection{Long-Time Limit}

Now, define a dynamical mapping $\Phi\mapsto\tilde\Phi$ similar to above by
$\Phi(t)=F^*+\tilde\Phi(t)$. Here, $F^*$ shall be the maximum fixed point of
${\mathcal I}$. It is easy to see that $\tilde\Phi$ fulfills the same
equations of motion as $\Phi$, provided one maps ${\mathcal F}\mapsto
\tilde{\mathcal F}$ as discussed above.

Since $\Phi$ is a completely monotone function, the limit
$\lim_{t\to\infty}\Phi(t)=G$ exists. The same also holds for $\tilde\Phi$, and
the mapping implies $G=F^*+\tilde G$, thus $G\succeq F^*$.
On the other hand, all time-derivatives of completely monotone functions
must vanish for long times, $\partial_t^n\Phi(t\to\infty)\to0$. Therefore,
one can integrate the time-domain equations of motion, Eq.~(\ref{eom}), to get
$S^{-1}G+N[G](G-S)=0$, which is equivalent to Eq.~(\ref{DW}). By this,
$G$ is a fixed point of Eq.~(\ref{DW}), and we have $G\preceq F^*$, from which
one concludes $G=F^*$.

Thus the maximum fixed point of Eq.~(\ref{DW}) corresponds to the
glass form factor matrix of the mixture.
In particular, we have explicitly generalized the iteration scheme of
Ref.~\cite{Goetze1995b}
that allows to calculate the form factors numerically
without solving the full equations of motion, Eq.~(\ref{eom}) or
Eq.~(\ref{time}). Let us also note that, with the above transformation,
$\tilde\Phi(t\to\infty)\to0$.

\section{Power Series Solutions}\label{powerseries}

\subsection{Power Series for Short Times}

In the time-domain, there exists a power series for $t<t_*$ with some
nonzero $t_*>0$,
\begin{equation}\label{tseries}
  \Phi(t)=\sum_{n=0}^\infty(-t)^n\phi_n\,,
\end{equation}
and analogous for $M(t)$, where $m_n=\sum_{k=0}^n{\mathcal F}[\phi_k,
\phi_{n-k}]$.

Let us suppose such a series exists. Then Eq.~(\ref{time}) implies
\begin{equation}
  \phi_{n+1}=\frac{1}{n+1}\left[\tau^{-1}S^{-1}\phi_n+\tau^{-1}\sum_{k=1}^n
  \frac{k!(n-k)!}{n!}m_{n-k}\phi_k\right]\,,
\end{equation}
for $n\ge1$, with $\phi_0=S$.
We prove by induction that
\begin{equation} 0\le t_*^n\|\phi_n\|\le\|\phi_0\| \end{equation}
for all $n$. Assuming the induction hypothesis, we have for $n+1$,
\begin{equation} t_*^{n+1}\|\phi_{n+1}\|\le\frac{t_*^n\|\phi_0\|}{n+1}
  \left[\|\tau^{-1}S^{-1}\|+\|\tau^{-1}\|\sum_{k=1}^n\frac{k!(n-k)!}{n!}
  \|m_{n-k}\|t_*^{n-k}\right]\,.
\end{equation}
Since the functional $\mathcal F$ is continuous, we can estimate $\|m_n\|\le
K\sum_{k=0}^n\|\phi_k\|\|\phi_{n-k}\|\le(n+1)K\|\phi_0\|$ with some constant
$K$. Furthermore, $\sum_{k=1}^n k!(n-k+1)!/(n+1)!\le\sum_{k=1}^n1/(n+1)\le1$,
and thus the proposition is proven for $t_*$ sufficiently small.

\subsection{Existence of Moments}

In the following, we restrict the discussion to regular points,
such that the spectral radius of the mapping $\psi$ is less than unity,
$r=1-2\varepsilon$ with some $\varepsilon>0$, say.
If one writes $S\tilde M(t)S=(\psi+\delta\psi[\tilde\Phi(t)])[\tilde\Phi(t)]$,
one has for
long times, $t>t_0$ with some fixed $t_0$, that $\delta\psi$ becomes arbitrarily
small as $\tilde\Phi\to0$,
$\delta\psi\preceq\varepsilon\psi$. Thus one infers an upper bound for
the memory kernel,
\begin{equation} S\tilde M(t)S\preceq
  (1-\varepsilon)\tilde\Phi(t)+\mu(t)\,,\quad t>0\,,
\label{SMSapprox}
\end{equation}
with some arbitrary
$\mu(t)\succeq0$ whose Laplace transform shall be analytic for small $|z|$.
For $\Imag z>0$ and $\Real z=0$,
the equivalent inequality holds for the Laplace-transformed
quantities, and inserting into Eq.~(\ref{eom}) yields
\begin{equation} 0\preceq-i\varepsilon\tilde\Phi(z)\preceq S\tau S-i\mu(z)
  \preceq S\tau S-i\mu(z=i0)\,.
\label{momentbound}
\end{equation}
This shows in particular that the zeroth moment of $\Phi(t)$,
$\Phi_0=\int_0^\infty\tilde\Phi(t)dt$ exists.
Due to Eq.~(\ref{momentbound}), we can, for any fixed $T$ and $\delta>0$, write
$\int_0^T\tilde\Phi(t)dt\preceq-i(1+\delta)\tilde\Phi(z=i/T\ln(1+\delta))
\preceq(1+\delta)(S\tau S-i\mu(z=i0))/\varepsilon$. Explicitly, we have then
\begin{equation} \Phi_0=S\tau S+SM_0S<\infty\,.\end{equation}
The existence of all other moments of $\tilde\Phi(t)$ and, from that, of
$\tilde M(t)$ shall be shown by induction. We denote these moments by
$\Phi_n$,
\begin{equation} \Phi_n:=\int_0^\infty t^n\tilde\Phi(t)dt\,,
\end{equation}
and equivalently for $\tilde M(t)$. To simplify notation, let us drop tildes
in the following.

First, we note that any $\Phi(t)\succeq0$ monotone and continuous such that
$\int_0^\infty t^n\Phi(t)dt$ is finite can be written for large times as
\begin{equation} \Phi(t)=\frac{K(t)}{t^{n+1}}\,,\quad t\to\infty\,,
\label{Kapprox}
\end{equation}
with some continuous $K(t)\succeq0$ satisfying $K(t)\to0$ for $t\to\infty$.
To see this, write for $t$ large, with $C$ some constant,
$0\preceq\Phi(t)t^{n+1}=C\Phi(t)\int_{t/2}^t\tau^nd\tau
\preceq C\int_{t/2}^\infty\tau^n\Phi(\tau)d\tau$, where the right-hand side
vanishes as $t\to\infty$.
\begin{subequations}
Since with $K(t)\succeq0$, also $\|K(t)\|$ and $Z(t)=\max_{T>t}\|K(T)\|$
are continuous, one can further estimate
\begin{equation} 0\le\|\Phi(t)\|\le\frac{Z(t)}{t^{n+1}}\,,\quad t\to\infty\,.
\label{Zapprox}
\end{equation}
For $n=0$, the bound can be improved. Write $t\Phi(t)\preceq
\Phi(t)/(\ln t-\ln T_0)\int_{T_0\ln T_0}^{t\ln t}d\tau\preceq
1/(\ln t-\ln T_0)\int_{T_0\ln T_0}^{t\ln t}\Phi(\tau)d\tau$, with some
$T_0$ satisfying $t>T_0\ln T_0>T_0$. From this, one gets
$\Phi(t)(t\ln t)\preceq K(t)\ln T_0+\int_{T_0\ln T_0}^{t\ln t}\Phi(\tau)d\tau
\preceq C$ for long times. One concludes $K(t)\preceq C/\ln t$, and
specializing to $T_0=\ln t$, one finds that the bound $C$ can be made
arbitrarily small, thus giving with obvious renaming
$\Phi(t)=\tilde K(t)/(t\ln t)$ with $\tilde K(t\to\infty)\to0$, or
\begin{equation} 0\le\|\Phi(t)\|\le\frac{\tilde Z(t)}{t\ln t}\,,\quad
  t\to\infty\,, \label{Zlogapprox}
\end{equation}
with continuous $\tilde Z(t)\succeq0$ vanishing as $t\to\infty$.
\end{subequations}

Using these expressions, one infers that the convolution $(M\ast\Phi)(t)$
decays to zero faster than $1/t^{n+1}$ for long times,
\begin{equation}\label{faltapprox}
(M\ast\Phi)(t)=\frac{\hat Z(t)}{t^{n+1}}\,,\quad t\to\infty\,,
\end{equation}
where $\hat Z(t)\to0$ for $t\to\infty$,
given that both $M$
and $\Phi$ have the above properties and finite $n$-th moments. Write
\begin{equation}
  \|t^{n+1}(M\ast\Phi)(t)\|\le\|t^{n+1}\int_0^{t/2}M(t')\Phi(t-t')dt'\|
  +\|t^{n+1}\int_0^{t/2}M(t-t')\Phi(t')dt'\|\,,
\end{equation}
and both terms on the right-hand side vanish for long times. To see this,
let us focus on the first term,
\begin{widetext}
\begin{equation}
  B(t):=t^{n+1}\|\int_0^{t/2}M(t')\Phi(t-t')dt'\|\le
  t^{n+1}\left(\max_{t/2\le T\le t}\|\Phi(T)\|\right)\left(C
  +\int_{T}^t\|M(t')\|dt'\right)\,,
\end{equation}
\end{widetext}
with some constant $C$. Now for $n\ge1$, we immediately have that
$\int_{T}^t\|M(t')\|dt'$ is bounded to above by some constant, and
Eq.~(\ref{Zapprox}) gives $B(t)\to\hat C\cdot Z(t)\to0$ for $t\to\infty$.
Similarly, for $n=0$ one uses the improved bound of Eq.~(\ref{Zlogapprox})
together with $\int_{T}^t\|M(t')\|dt'\le C/t$ to get $B(t)\to\tilde C
\tilde Z(t)/\ln t\,(C'+\ln t)\to0$ as $t\to\infty$, which completes the proof.

Now, suppose all moments of $\Phi(t)$ and $M(t)$ up to $n$ exist. It is clear,
that then also the moments of $(M\ast\Phi)(t)$ up to $n$ exist.
From Eqs.~(\ref{time}) and (\ref{SMSapprox}), we get
$\varepsilon\Phi(t)\preceq\mu(t)-\frac{d}{dt}[S(M\ast\Phi)(t)-S\tau\Phi(t)]$,
which can be integrated to give, together with Eq.~(\ref{faltapprox}),
\begin{equation}\label{momentapprox}
  \varepsilon\Phi_{n+1}\preceq\mu_{n+1}+(n+1)S(M\ast\Phi)_n-(n+1)S\tau\Phi_n\,.
\end{equation}
This implies the existence of both the $(n+1)$-th moment of $\Phi(t)$ and
of $M(t)$.

\subsection{Power Series for Small Frequencies}

Having proven the existence of all moments, we can proceed to establish a
power series for $\tilde\Phi(z)$ and $\tilde M(z)$ for small $|z|$.
We show that the series
\begin{equation}\label{zseries}
  \Phi(z)+F^*/z=\sum_{n=0}^\infty\frac{i\Phi_n\,(iz)^n}{n!}
\end{equation}
converges for $|z|<z_0$, with $z_0>0$ a non-zero radius of convergence.

For the $n$-th term in the power series, Eq.~(\ref{zseries}), there holds
the inequality
\begin{equation} \frac{z_0^n\|\Phi_n\|}{n!}\le\frac{\|\Phi_0\|}{(n+1)^2}\,,
\quad n\ge1\,,
\end{equation}
which can be proven by induction.
Taking norms on both sides of Eq.~(\ref{momentapprox}) one obtains,
together with $\|SM_kS\|\le\|\Phi_k\|+\|\mu_k\|$,
\begin{equation}
  \varepsilon\|\Phi_{n+1}\|\le\|\mu_{n+1}\|+(n+1)\sum_{k=0}^n\left(
  \|\Phi_k\|+\|\mu_k\|\right)\|S^{-1}\|\|\Phi_{n-k}\|+(n+1)\|S\tau\|
  \|\Phi_n\|\,.
\end{equation}
But $\mu(z)$ is analytic for small $|z|$, so let us estimate
$\mu_n\le Ct_0^nn!$ with some constant $C>0$ and some $t_0>0$. We also
choose $z_0$ such that $z_0t_0<1/2$, in order to get from the induction
hypothesis
\begin{widetext}
\begin{subequations}
\begin{align}
  \varepsilon\frac{z_0^{n+1}\|\Phi_{n+1}\|}{(n+1)!}
  &\le C(z_0t_0)^{n+1}+\frac{z_0\|\Phi_0\|}{(n+1)^2}
  \left[\|S^{-1}\|\sum_{k=0}^n\left(C(z_0t_0)^k+\frac{\|\Phi_0\|}{(k+1)^2}
  \right)+\|S\tau\|\right]\\
  &\le \frac{5C(z_0t_0)}{(n+2)^2}
  +\frac{9z_0\|\Phi_0\|}{(n+2)^2}\left[\|S^{-1}\|\left(2C+\|\Phi_0\|\frac{\pi^2}
  {6}\right)+\|S\tau\|\right]\le\frac{\varepsilon\|\Phi_0\|}{(n+2)^2}
\,,
\end{align}
\end{subequations}
\end{widetext}
where we have used $2^{-n}<5/(n+2)^2$ and $1/(n+1)^2<9/(n+2)^2$ and the
last inequality holds for small enough $z_0$. From this, the 
convergence of the power series of Eq.~(\ref{zseries}) follows. Consider
Cauchy sequences for $m\ge n$, $|z|<z_0$,
\begin{equation}
  \|\sum_{k=n}^m\frac{i\Phi_k\,(iz)^k}{k!}\|\le\sum_{k=n}^\infty
  \frac{\|\Phi_0\|}{(k+1)^2}\,,
\end{equation}
where the right-hand side becomes arbitrarily small if $n$ becomes large.

\section{Conclusion}\label{conclusio}

We have shown that, for a multi-component colloidal mixture
driven by Smoluchowski dynamics, the mode-coupling theory
of the glass transition (MCT) provides an approximation to the density
correlators that preserves the complete monotonicity imposed by the
general structure of the time-evolution operator. Thus, positivity of the
spectra is guaranteed. Since the correlation functions are superpositions of
purely decaying exponentials in the sense of Eq.~(\ref{represent}), the term
``structural relaxation'' given to the dynamics close to a glass transition
is justified. The glass form factor was found to be determined
by a maximum principle. It can be evaluated by an iteration scheme whose
linearization fulfills the prepositions needed for a generalized
Perron-Frobenius theorem under the natural assumption that the system has
no decoupling wave-vector subspaces.
This in turn ensures a non-degenerate maximum
eigenvalue that was shown to be smaller than, or equal to unity for all
physical states. Thus within the MCT of multi-component systems, the
only possible glass-transition singularities are bifurcations of type $A_\ell$,
occuring at points where the Perron-Frobenius eigenvalue equals unity.

The proofs presented here suggest numerical schemes for the solution of the
nonlinear equations of motion, that are stable in the sense that they will
converge to the uniquely defined, symmetric, completely monotone solution.
In particular,
the glass form factor can be found without solving the dynamical equations.

\begin{figure}
\includegraphics[width=.9\linewidth]{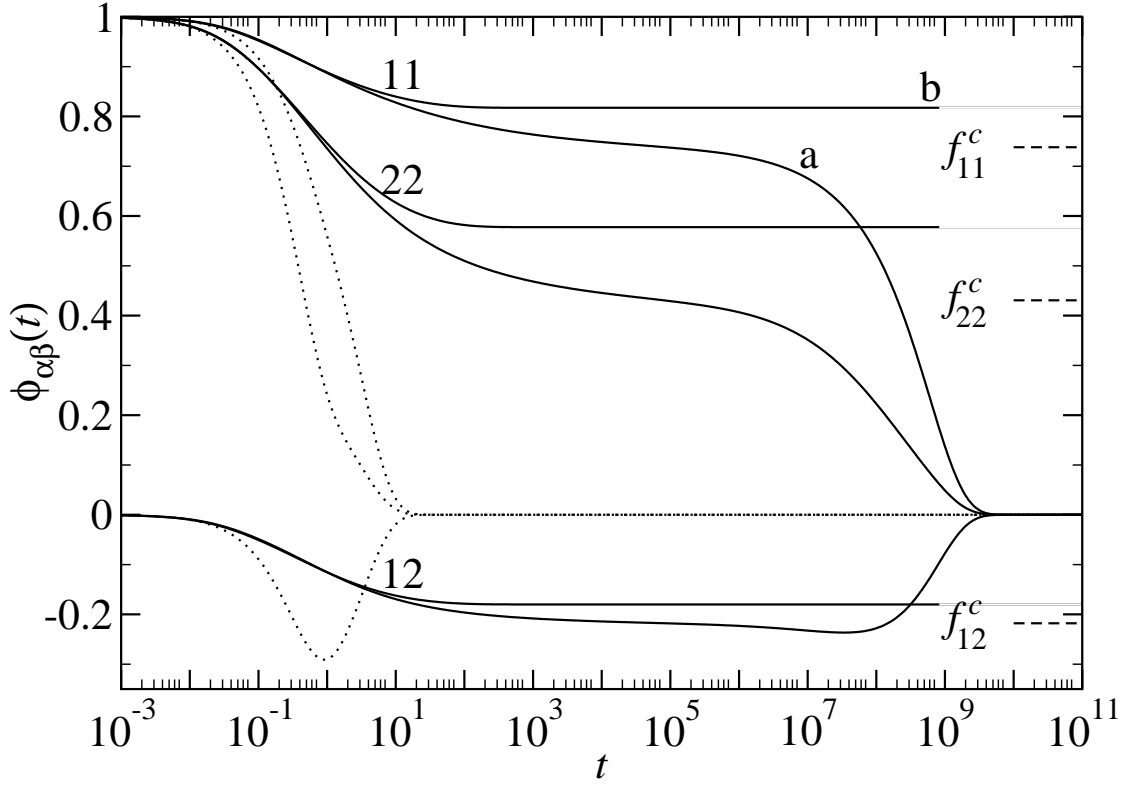}
\caption{\label{fig1}Normalized correlation functions
$\phi_{\alpha\beta}(q,t)=\left(S^{-1/2}(q)\Phi(q,t)S^{-1/2}(q)\right)
_{\alpha\beta}$ for a binary hard-sphere mixture with diameter ratio
$d_{\text{B}}/d_{\text{A}}=0.6$, packing fraction of $\text{B}$-particles
$\varphi_{\text{B}}/\varphi=0.2$, wave vector $q=6.2/d_{\text{A}}$,
calculated from Eq.~(\protect\ref{eom}); details see text. The labels
$\text{11}$, $\text{12}$, and $\text{22}$ indicate the three independent
matrix elements, and labels $\text{a}$ and $\text{b}$ correspond to total
packing fractions $\varphi=0.5195$ and $0.525$, respectively. Here,
$\varphi=(\pi N/6V)(x_{\text{A}}d_{\text{A}}^3+x_{\text{B}}d_{\text{B}}^3)$,
with $x_\alpha=N_\alpha/N$ being the number concentrations.
The dashed horizontal lines indicate the long-time limit for
$\varphi=\varphi^c\approx0.519608$, and the dotted curves correspond to an
exponential decay as described in the text.
}
\end{figure}

Typical solutions to the MCT equations appear in Fig.~\ref{fig1}.
We chose as the simplest case a binary mixture of hard spheres. Then the
state of the system is determined by three numbers, which we take to be
the total packing fraction, $\varphi=\varphi_{\text{A}}+\varphi_{\text{B}}$,
the diameter ratio $\delta=d_{\text{B}}/d_{\text{A}}$, and the packing
contribution of the second species, $\varphi_{\text{B}}/\varphi$.
Here, the species are labeled with
subscripts $\text{A}$ and $\text{B}$, and $\varphi_\alpha=(\pi d_\alpha^3/6)
(N_\alpha/V)$ are the partial volumes occupied by each species.
To calculate the MCT vertex, Eq.~(\ref{vertex}), one needs to know the
static structure factor of the system, as well as static
three-particle correlation functions. For the latter, to our knowledge,
no analytic expressions are available, and thus we follow the commonly applied
approximation through two-particle correlations \cite{Fuchs1993}. The former
shall be approximated
using the well-known Percus-Yevick (PY) approximation to the Ornstein-Zernike
integral equation. Within this framework, ``PY-exact'' solutions are
available \cite{Lebowitz1964b}. As was already mentioned in connection
with Eq.~(\ref{vertex}), the approximations involved in evaluating the vertex
are of no importance for the mathematical aspects of the solutions we wish
to demonstrate.

Figure~\ref{fig1} shows normalized correlation functions
$\phi(q,t)=S^{-1/2}(q)\Phi(q,t)S^{-1/2}(q)$ for a particular wave vector
$q=6.2/d_{\text{A}}$. They
are numerical solutions of Eq.~(\ref{eom}) with $\tau_{\alpha\beta}(q)=1/(q^2
D^0_\alpha)\delta_{\alpha\beta}$ and $D^0_\alpha=0.005/R_\alpha$, on a grid
of 100 wave vectors $q=0.2,\dots 39.8$ with $\Delta q=0.4$. The
size ratio and the concentrations are kept fixed at $(\delta,\varphi_{\text{B}}
/\varphi)=(0.6,0.2)$, and the total packing fraction is varied as
indicated in the figure caption;
a procedure corresponding to what has been done in experiment
\cite{Williams2001b}.
Note that while the two diagonal matrix elements are positive and monotonically
decreasing for all times $t$, both needs not be true in general for the
off-diagonal elements.
For comparison also shown is the suggested starting point of
Eq.~(\ref{laplace-iteration}),
$\phi^{(0)}(t)=S^{-1/2}\exp[-(S\tau)^{-1}t]S^{1/2}$, which is completely
monotone by construction. One notices a drastic slowing down in the relaxation
of the correlation functions towards their long-time limits.
This demonstrates the dynamics close to a critical point where the
Perron-Frobenius eigenvalue $r=1$, which in this system occurs at
$\varphi^c\approx0.519608$ and which is the concern of the asymptotic solutions
to mode-coupling theory. The solutions for the long-time limit at
$\varphi=\varphi^c$ as determined from Eq.~(\ref{DW}) are included in the
figure.

\begin{figure}
\includegraphics[width=.9\linewidth]{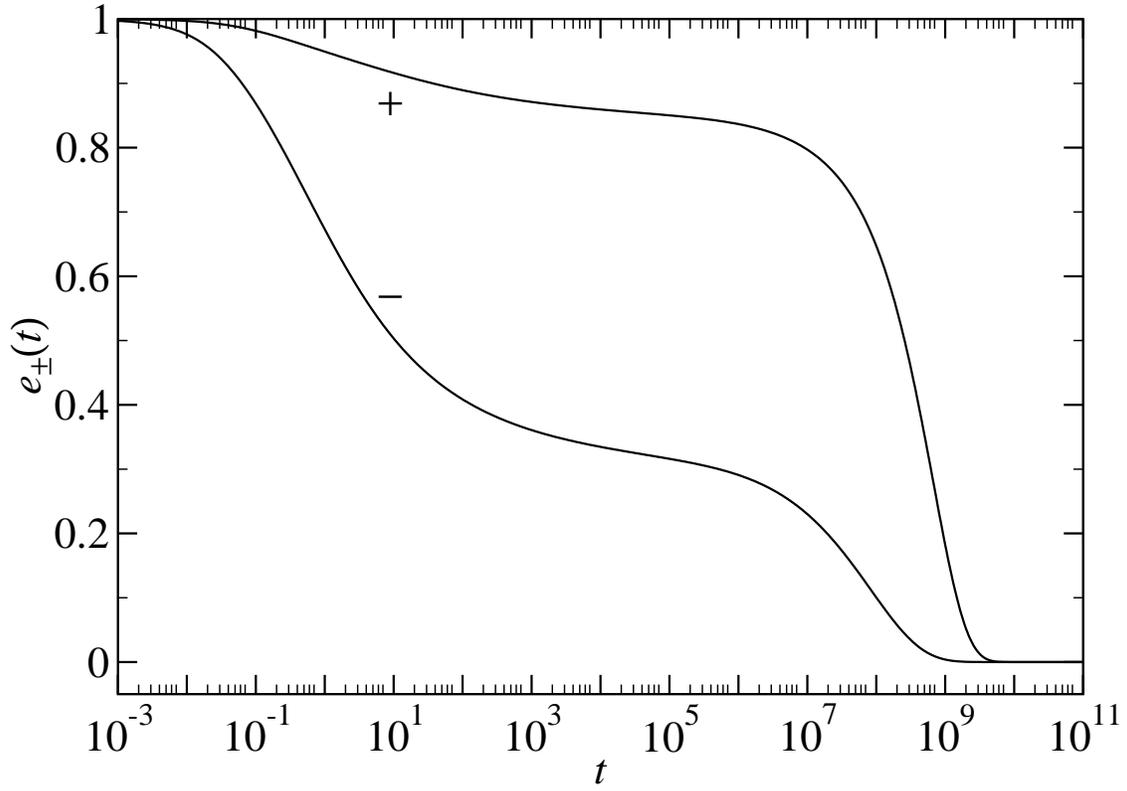}
\caption{\label{fig2}Eigenvalues $e_\pm(q,t)$ of $\phi(q,t)$ from
Fig.~\protect\ref{fig1} for $\varphi=0.5195$ (all other parameters as given
in Fig.~\protect\ref{fig1}) as functions of time $t$.}
\end{figure}

Complete monotonicity requires all eigenvalues
of $\phi(t)$ to be positive for all $t$. To demonstrate that our numerical
solution is in accordance with this, we show in Fig.~\ref{fig2} the
eigenvalues $e_\pm(t)$ as functions of time for a fixed total packing
fraction. One clearly recognizes the above statement to hold.

A power series in the time domain exists, and at regular points, i.e.\ for
vertices such that the Perron-Frobenius eigenvalue is smaller than unity,
also for small frequencies the power series has a nonzero radius of
convergence.
In particular, the existence of all moments of the density correlation
function at regular points, and of a finite radius of convergence for the
frequency-domain power series implies the existence of a final exponential
relaxation,
\begin{equation} \Phi(t)-F^*={\mathcal O}(e^{-\gamma_0t})\,.
\end{equation}
This holds, since Eq.~(\ref{zseries}) implies that the measure $a$ in
Eq.~(\ref{represent}) has an atom of mass $F^*$ at $\gamma=0$, is constant
for $0<\gamma<\gamma_0$, and has a point of increase at $\gamma=\gamma_0$
\cite{Feller1971b}. Thus,
\begin{equation} \Phi_{\mu\nu}(q,t)=F_{\mu\nu}^*(q)
  +\int_{\gamma_0}^\infty e^{-\gamma t}\,da_{\mu\nu,q}(\gamma)\,.
\end{equation}

The mode-coupling theory for mixtures can also be applied to systems with
Newtonian, instead of stochastic, short-time dynamics, e.g.\ metallic melts.
Furthermore, a recent extension of MCT to molecular liquids that treats each
molecule as consisting of $s$ constituent sites also leads to substantially
the same equations \cite{Chong2002}.
Since in these cases, the representation of Eq.~(\ref{represent}) through
decaying exponentials only will not be valid in general, the proofs
presented here cannot readily be applied. Work published for one-component
Newtonian systems \cite{Haussmann1990} suggests that existence
and uniqueness of the solution, even if it will
not be completely monotone, can nevertheless be proven. This has, however,
not been done so far.

It is an observation of both theory \cite{Franosch1998} and
computer experiment \cite{Gleim1998}, that the different short-time behavior
does not influence the dynamics at sufficiently long times apart from an
overall shift in time scale, given a strong enough coupling such that MCT
contributions are important.
One then expects the long-time limit of the Newtonian dynamics solutions to
exist and to be
governed by Eq.~(\ref{DW}). Note that the properties of this equation and
its maximum fixed point do not depend on the short-time dynamics, nor do the
commonly applied asymptotic formulas.
Similarly, the prediction of only $A_\ell$ singularities as glass transitions
will remain valid as long as the linearization of Eq.~(\ref{DW}) is
irreducible. This can be expected unless some special symmetry will introduce
zero-couplings in the vertex, which in principle can happen within the
molecular site-site description of Ref.~\cite{Chong2002}.

\begin{acknowledgments}
We acknowledge $\ldots$ [to be completed]
Financial support was provided through DFG grant No.~Go.154/12-1.
\end{acknowledgments}

\begin{appendix}
\section{Perron Theorem}\label{perron}
Let us sketch here for completeness some results generalized from the
Perron-Frobenius theorem for irreducible matrices \cite{Gantmacher1974b}.
For a generalization of the complete theorem for positive linear maps on
$C^\star$ algebras, the reader is referred to Ref.~\cite{Evans1978}.

As above, let ${\mathcal A}^M$ denote the $C^\star$ algebra of $M$-vectors
of $s\times s$ matrices over $\complex$.
Consider the positive linear map $\psi$, which
maps the set of symmetric, real, positive definite elements ${\mathcal A}^M_+$
onto itself, $\psi[a]\succeq0$ for $a\succeq0$.
$\psi$ is called \emph{irreducible}, if there exists some positive, finite
number $n$ such that $T[a]\succ0$ for $a\succeq0$ and
\begin{equation} T[a]=(1+\psi)^n[a]\,.\end{equation}
If $\psi$ is irreducible, we have that $\psi[a]\succ0$ if $a\succ0$.

Now, define a mapping $r:({\mathcal A}^M_+,\complex^s)\to\real$ by
\begin{equation} r(a,v)=\min_{1\le q\le M}\frac{(v|\psi_q[a]v)}
  {(v|a_qv)}\,,
\end{equation}
where $q$ labels the elements of $a\in{\mathcal A}^M$, $a=(a_q)_{q=1,\dots M}$,
and $(\cdot|\cdot)$ is a scalar product over $\complex^s$.
Furthermore, set $r(a)=\inf_{v\in\complex^s}r(a,v)=\inf_{v\in\sphere^s}r(a,v)$,
where $\sphere^s$ denotes the $s$-dimensional unit sphere, and the latter
equation holds since $r(a,\lambda v)$ with $\lambda\in\complex$ is independent
of $\lambda$.

One immediately checks $r(a)\ge0$ and
\begin{equation}\label{frmax}
  \psi[a]\succeq r(a)a\,.
\end{equation}
However, $r(a,v)$ is not continuous on $({\mathcal A}^M_+,\complex^s)$.
Let us define a set
\begin{equation}
  {\mathcal B}:=\{b;b=T[a],a\in{\mathcal A}^M_+,\|a\|=1\}\,.
\end{equation}
Then, $b\succ0$ for any $b\in{\mathcal B}\subset{\mathcal A}^M_+$.
Since $r(b,v)$ is continuous on the closed and compact set
$(\mathcal B,\sphere^s)$, it assumes its minimum with respect to $v$.
It follows that on $\mathcal B$, $r(b)$ fulfills a maximum principle in the
sense that it is the maximum $\real$-number for which $\psi[b]\succeq r(b)b$.
Furthermore, for $a\in{\mathcal A}^M_+$ and $b=T[a]\in{\mathcal B}$
we have $T[\psi(a)-r(a)a]=\psi[b]-r(a)b\succ0$, and by the maximum principle
we get $r(b)\ge r(a)$.

Next define
\begin{equation} r=\sup_{a\in{\mathcal A}^M_+}r(a)\,.\end{equation}
Since $r(1)>0$, clearly $r>0$.
The supremum can be restricted to $b\in\mathcal B$ since the above
inequality holds. But there, $r(b)$ assumes its maximum, and thus $r$
attains the supremum for some extremal vector $z\succ0$.

We continue by showing that $r$ indeed is an eigenvalue of $\psi$ and equal
to the spectral radius, and that the corresponding eigenvector $z$ is unique,
i.e.\ the eigenvalue is non-degenerate.

Assume, $\psi[z]-rz\succeq0$ but not the null element. Then $\psi[\hat z]
-r\hat z\succ0$ with $\hat z=T[z]$, but the maximum principle then implies
$r(\hat z)>r$ in contradiction to the definition of $r$. Thus, $r$ is an
eigenvalue of $\psi$. Suppose now, there are two eigenvectors $z$, $z'$
corresponding to this eigenvalue. We then can find some $\lambda\in\real$
such that $\lambda z-z'\succeq0$ but not strictly positive definite. But this
implies $T[\lambda z-z']=(1+r)^n(\lambda z-z')\succ0$, in contradiction to the
construction of $\lambda$. Thus, the eigenvalue $r$ is non-degenerate.

Now for any $a\in{\mathcal A}^M_+$, define the mapping
\begin{equation} \sigma[a]=(1/r)z^{-1/2}\psi[z^{1/2}az^{1/2}]z^{-1/2}\,.
\end{equation}
Since $\sigma[1]=1$, $\|\sigma\|=1$. Suppose $\psi[u]=\alpha u$ with some
$\alpha\in\complex$. Write $v=z^{-1/2}uz^{-1/2}$, which gives $\sigma[v]
=(\alpha/r)v$.  But for any eigenvalue $\lambda$ of
$\sigma$, we have $|\lambda|\le\|\sigma\|$, and thus in particular
$|\alpha|\le r$. Therefore, $r$ is the spectral radius of $\psi$.

\end{appendix}

\bibliography{mct}
\bibliographystyle{apsrev}

\end{document}